\documentclass[aps,11pt]{revtex4}
\usepackage[dvips]{graphicx}
\usepackage{amsmath}
\usepackage{bm}

\begin{document}

\preprint{}

\title{UNIFICATION WITHOUT SUPERSYMMETRY:\\
 NEUTRINO MASS, PROTON DECAY AND LIGHT LEPTOQUARKS}
\author{Ilja Dorsner}
\email{idorsner@ictp.trieste.it}
\affiliation{The Abdus Salam
International Centre for Theoretical Physics\\
Strada Costiera 11, 31014 Trieste, Italy.}
\author{Pavel Fileviez P\'erez}
\email{fileviez@higgs.fis.puc.cl, filevi@mppmu.mpg.de}
\affiliation{Pontificia Universidad Cat\'olica de Chile, Facultad de
 F{\'\i}sica\\
Casilla 306, Santiago 22, Chile.}


\begin{abstract}
We investigate predictions of a minimal realistic
non-supersymmetric $SU(5)$ grand unified theory. To accomplish
unification and generate neutrino mass we introduce one extra
Higgs representation---a $\bm{15}$ of $SU(5)$---to the particle
content of the minimal Georgi-Glashow scenario. Generic prediction
of this setup is a set of rather light scalar leptoquarks. In the
case of the most natural implementation of the type II see-saw
mechanism their mass is in the phenomenologically interesting
region ($O(10^2$--$10^3)$\,GeV). As such, our scenario has a
potential to be tested at the next generation of collider
experiments, particularly at the Large Hadron Collider (LHC) at
CERN. The presence of the $\bm{15}$ generates additional
contributions to proton decay which, for light scalar leptoquarks,
can be more important than the usual gauge $d=6$ ones. We
exhaustively study both and show that the scenario is not excluded
by current experimental bounds on nucleon lifetimes.
\end{abstract}

\maketitle

\section{Introduction}
The most predictive grand unified theory (GUT) based on an $SU(5)$
gauge symmetry is a minimal non-supersymmetric model of Georgi and
Glashow~\cite{GG} (GG). However, the failure to accommodate
experimentally observed fermion masses and mixing and to unify
electroweak and strong forces decisively rule it out.
Nevertheless, the main features of the underlying theory, e.g,
partial matter unification and one-step symmetry breaking, are so
appealing that there has been a number of proposals to enlarge its
structure by adding more representations to have the theory in
agreement with experimental data. Since the number of possible
extensions is large it is important to answer the following
question. What is the minimal number of extra particles that
renders an $SU(5)$ gauge theory realistic? Such an extension of
the GG model with the smallest possible number of extra particles
that furnish full $SU(5)$ representation(s) has all prerequisites
to be the most predictive one. Thus, if there is a definite answer
to the first question it is important to ask the second one: What
are the possible experimental signatures and associated
uncertainties of such a minimal setup? If uncertainties of the
minimal extension are significant the same is even more true of a
more complicated structure unless additional assumptions are
imposed. We address both questions in great detail and present
truly minimal, i.e., minimal in terms of number of fields,
realistic non-supersymmetric $SU(5)$ scenario.

As we demonstrate later, unification, in the minimal scenario,
points towards existence of light scalar leptoquarks which could
generate very rich phenomenological signatures. These are looked
for in the direct search experiments as well as in the experiments
looking for rare processes. Their presence generates novel proton
decay contributions which can be very important. Moreover, since
in our scheme their coupling to matter is through the Majorana
neutrino Yukawas, their observation might even allow measurements
of and/or provide constraints on neutrino Yukawa coupling entries.
This makes our scenario extremely attractive. At the same time,
rather low scale of vector leptoquarks that is inherent in
non-supersymmetric theories exposes our scenario to the tests via
nucleon lifetime measurements. We investigate all relevant
experimental signatures of the scenario, including its status with
respect to the present bounds on nucleon lifetime.

In the next section we define our framework. In Section~\ref{III}
we discuss how it is possible to get gauge couplings unification
in agreement with low energy data. Then, in Section~\ref{IV} we
discuss possible experimental signatures of the minimal scenario.
We conclude in Section~\ref{conclusions}. Appendix~\ref{first}
contains relevant details and notation of the minimal
non-supersymmetric realistic $SU(5)$ we refer to throughout the
manuscript. The origin of theoretical bounds on the familiar gauge
$d=6$ proton decay operators is critically analyzed in
Appendix~\ref{second}. Appendix~\ref{third} contains details on
the two-loop running of the gauge couplings that is presented for
completeness of our work.
\section{A minimal realistic non-supersymmetric $SU(5)$ scenario}
\label{SU(5)theory} In order to motivate the minimal $SU(5)$ grand
unifying theory (GUT), where we define such a theory to be the one
with the smallest possible particle content that renders it
realistic, we first revisit the GG model~\cite{GG} and discuss its
shortcomings. Only then do we present the minimal realistic
scenario and investigate its experimental signatures and related
uncertainties.

The GG model fails from the phenomenological point of view for a
number of reasons:
\begin{enumerate}
    \item It does not incorporate massive neutrinos;
    \item It yields charged fermion mass ratios in gross violation
    of experimentally observed values;
    \item It cannot account for the gauge coupling
unification. (For one of the first rulings on (non)unification see
for example Ref.~\cite{Marciano}.)
\end{enumerate}

The first flaw is easy to fix; one either introduces three
right-handed neutrinos---singlets of the Standard Model (SM)---to
use a type I see-saw mechanism~\cite{seesaw} to generate their
mass or adds a Higgs field---a $\bm{15}$ of $SU(5)$---to generate
neutrino mass through the so-called type II
see-saw~\cite{Lazarides,Mohapatra}. One might also use the
combination of the two. The fourth option to use the Planck mass
suppressed higher-dimensional operators~\cite{Weinberg,Barbieri}
does not look promising since it generates too small scale for
neutrino mass to explain the atmospheric and solar neutrino data.
Nevertheless, it might still play an important
role~\cite{GoranBerezhiani}. We focus our attention on the second
option, i.e., addition of $\bm{15}$, in view of the fact that the
right-handed neutrinos, being singlets of the SM, do not
contribute to the running. Hence, their mass scale cannot be
sufficiently well determined or constrained unless additional
assumptions are introduced.

The second flaw can be fixed by either introducing the
higher-dimensional operators in the Yukawa sector~\cite{Ellis:1}
or resorting to a more complicated Higgs sector {\it \`a la}\/
Georgi and Jarlskog~\cite{GJ}. The former approach introduces a
lot more parameters into the model (see for example~\cite{Bajc2})
but unlike in the neutrino case these operators might have just a
right strength to modify ``bad" mass predictions for the charged
fermions. In order to keep as minimal as possible the number of
particles we opt for the scenario with the non-renormalizable
terms.

The third flaw requires presence of additional non-trivial split
representations besides those of the GG $SU(5)$ model. It can thus
be fixed in conjunction with the first and second one. For
example, introduction of an extra $\bm{45}$ of Higgs to fix mass
ratios of charged fermions {\it \`a la}\/ Georgi and
Jarlskog~\cite{GJ} allows one to achieve unification and, at the
same time, raise the scale relevant for proton decay. (For the
studies on the influence of an extra $\bm{45}$ on the running and
other predictions see for example~\cite{BabuMa,Giveon}.) We will
see that the addition of one extra $\bm{15}$ of Higgs plays a
crucial role in achieving the unification in our case.

So, what we have in mind as the minimal realistic $SU(5)$ model is
the GG model supplemented by the $\bm{15}$ of Higgs to generate
neutrino mass and which incorporates non-renormalizable effects to
fix the Yukawa sector of charged fermions. We analyze exclusively
non-supersymmetric scenario for the following three reasons:
first, this guarantees the minimality of the number of fields;
second, there are no problems with the $d=4$ and $d=5$ proton
decay operators; third, since the grand unifying scale is lower
than in supersymmetric scenario the setup could possibly be
verified or excluded in the next generation of proton decay
experiments.
\section{Unification of gauge couplings}
\label{III} The main prediction, besides the proton decay, of any
GUT is the unification of the strong and electroweak forces. We
thus show that it is possible to achieve gauge coupling
unification in a consistent way in our scenario.

At the one-loop level the running of gauge couplings is given by
\begin{equation}
\label{running}
\alpha^{-1}_i|_{M_Z}=\alpha^{-1}_{GUT}+\frac{1}{2
\pi} b_i \ln \frac{M_{GUT}}{M_Z},
\end{equation}
where $i=1,2,3$ for $U(1)$, $SU(2)$, and $SU(3)$, respectively.
$b_i$ are the appropriate one-loop coefficients~\cite{Cheng} and
$\alpha_{GUT}=g^2_{GUT}/(4 \pi)$ represents the gauge coupling at
the unifying scale $M_{GUT}$. The SM coefficients for the case of
$n$ light Higgs doublet fields are:
\begin{equation}
b_1=\frac{40}{10}+\frac{n}{10},
\qquad\qquad
b_2=-\frac{20}{6}+\frac{n}{6},
\qquad\qquad
b_3=-7.
\end{equation}
Even though the SM coefficients do not generate unification in
both the $n=1$ and $n=2$ case for any value of $\alpha_{GUT}$ and
$M_{GUT}$ that is not an issue since the SM does not predict the
gauge coupling unification in the first place. On the other hand,
a GUT, which does predict one, automatically introduces a number
of additional particles with respect to the SM case that, if light
enough, can change the outcome of the SM running. This change is
easily incorporated if one replaces $b_i$ in Eq.~\eqref{running}
with the effective one-loop coefficients $B_i$ defined by
\begin{equation}
\label{r}
B_i = b_i+\sum_{I} b_{iI} r_{I}, \qquad r_I=\frac{\ln
M_{GUT}/M_{I}}{\ln M_{GUT}/M_{Z}},
\end{equation}
where $b_{iI}$ are the one-loop coefficients of any additional
particle $I$ of mass $M_I$ ($M_Z \leq M_I \leq M_{GUT}$).
Basically, given a particle content of the GUT and
Eqs.~\eqref{running} and~\eqref{r} we can investigate if the
unification is possible.

Following Giveon {\it et al}~\cite{Giveon}, Eqs.~\eqref{running}
can be further rewritten in a more suitable form in terms of
differences in the effective coefficients $B_{ij}(=B_i-B_j)$ and
low energy observables. They find two relations that hold at
$M_Z$:
\begin{subequations}
\begin{eqnarray}
\frac{B_{23}}{B_{12}}&=&\frac{5}{8} \frac{\sin^2
\theta_w-\alpha_{em}/\alpha_s}{3/8-\sin^2 \theta_w},\\
\nonumber\\
\ln
\frac{M_{GUT}}{M_Z}&=&\frac{16 \pi}{5 \alpha_{em}}
\frac{3/8-\sin^2 \theta_w}{B_{12}}.
\end{eqnarray}
\end{subequations}
Adopting the following experimental values at $M_Z$ in the
$\overline{MS}$ scheme~\cite{PDG2004}: $\sin^2 \theta_w=0.23120
\pm 0.00015$, $\alpha_{em}^{-1}=127.906 \pm 0.019$ and
$\alpha_{s}=0.1187 \pm 0.002$, we obtain
\begin{subequations}
\label{conditions}
\begin{eqnarray}
\label{condition1}
\frac{B_{23}}{B_{12}}&=&0.719\pm0.005,\\
\nonumber\\
\label{condition2}
\ln \frac{M_{GUT}}{M_Z}&=&\frac{184.9 \pm 0.2}{B_{12}}.
\end{eqnarray}
\end{subequations}
Last two equations allow us to constrain the mass spectrum of
additional particles that leads to an exact unification at
$M_{GUT}$. (In what follows we consistently use central values
presented in Eqs.~\eqref{conditions} unless specified otherwise.
The inclusion of the two-loop effects and threshold corrections is
addressed in detail in Appendix~\ref{third}.)

The fact that the SM with one (two) Higgs doublet(s) cannot yield
unification is now more transparent in light of
Eq.~\eqref{condition1}. Namely, the resulting SM ratio is simply
too small ($B_{23}/B_{12}=0.53$ for $n=1$) to satisfy equality in
Eq.~\eqref{condition1}. What is needed is one or more particles
that are relatively light and with suitable $b_i$ coefficients
that can increase the value of the $B_{23}/B_{12}$ ratio. The most
efficient enhancement is realized by a field that increases
$B_{23}$ and decreases $B_{12}$ simultaneously. For example, light
Higgs doublet is such a field (see the $\Psi_D(\subset \bm{5}_H)$
coefficients in Table~\ref{tab:table2}) and it takes at least
eight of them, at the one-loop level, to bring $B_{23}/B_{12}$ in
accord with experiments. Other fields that could generate the same
type of improvement in our scenario are light $\Sigma_3(\subset
\bm{24}_H)$, $\Phi_a(\subset \bm{15}_H)$ and $\Phi_b(\subset
\bm{15}_H)$. $B_{ij}$ coefficients of all the particles in our
scenario are presented in Table~\ref{tab:table2} and the relevant
notation is set in Appendix~\ref{first}.
\begin{table}[h]
\caption{\label{tab:table2} Contributions to the
$B_{ij}$coefficients. The mass of the SM Higgs doublet is taken to
be at $M_Z$.}
\begin{ruledtabular}
\begin{tabular}{lccccccccc}
     &Higgsless SM&$\Psi_D$&$\Psi_T$ & $V$ & $\Sigma_8$
     & $\Sigma_3$ & $\Phi_a$ & $\Phi_b$ & $\Phi_c$\\
\hline $B_{23}$& $\frac{11}{3}$&$\frac{1}{6}$&$-\frac{1}{6}
r_{\Psi_T}$ &$-\frac{7}{2}r_V$ &$-\frac{1}{2}
r_{\Sigma_8}$&$\frac{1}{3} r_{\Sigma_3}$ &$\frac{2}{3}r_{\Phi_a}$
&$\frac{1}{6} r_{\Phi_b}$ &$-\frac{5}{6} r_{\Phi_c}$\\
$B_{12}$&$\frac{22}{3}$&$-\frac{1}{15}$&$\frac{1}{15} r_{\Psi_T}$
&$-7r_V$ &0 &$-\frac{1}{3} r_{\Sigma_3}$
&$-\frac{1}{15}r_{\Phi_a}$ &$-\frac{7}{15} r_{\Phi_b}$ &$\frac{8}{15} r_{\Phi_c}$\\
\end{tabular}
\end{ruledtabular}
\end{table}

The improvement can also be due to the field that lowers $B_{12}$
only or lowers $B_{12}$ at sufficiently faster rate than $B_{23}$.
Looking at Table~\ref{tab:table2} we see that the superheavy gauge
fields $V$ comprising $X$ and $Y$ gauge bosons and their conjugate
partners can accomplish the latter. (Note that the gauge
contribution improves unification at the one-loop level only if
$n\neq0$ and the improvement grows with the increase of
$n$~\cite{Giveon,MY}. This is because the $B_{23}/B_{12}$ ratio of
the Higgsless SM coefficients is the same as for the corresponding
ratio of $V$ coefficients.) But, their contribution to running has
to be subdominant; otherwise one runs into conflict with the
experimental data on nucleon lifetimes.

All in all, the fields capable of improving unification in our
minimal $SU(5)$ grand unified scenario are $\Psi_D$, $\Sigma_3$,
$\Phi_a$, $\Phi_b$ and $V$. Again, we refer reader to
Appendix~\ref{first} for our notation. We treat their masses as
free parameters and investigate the possibility for consistent
scenario with the exact one-loop unification. Since all other
fields in the Higgs sector, i.e., $\Psi_T$, $\Sigma_8$ and
$\Phi_c$, simply worsen unification we simply assume they live at
or above the grandunifying scale.

In order to present consistent analysis we now discuss the
constraints coming from proton decay on $B_{ij}$ coefficients.
These enter via Eq.~\eqref{condition2} and assumption that
$M_V=M_{GUT}$. As we show these constraints are rather weak if the
gauge $d=6$ contributions are dominant as is usually assumed in
non-supersymmetric GUTs~\cite{Buras}. For example, if we use the
latest bounds on nucleon decay lifetimes we obtain, in the context
of an $SU(5)$ non-supersymmetric GUT, in the case of maximal
(minimal) suppression in the Yukawa sector~\cite{Ilja3} $M_V>2.5
\times 10^{13}$\,GeV ($M_V>1.5 \times 10^{15}$\,GeV). (The minimal
suppression case corresponds to the GG scenario with $Y_U=Y^T_U$
and $Y_D=Y^T_E$, where $Y_U$, $Y_D$ and $Y_E$ are the Yukawa
matrices of charged fermions. Non-renormalizable contributions
violate both of those relations. The same is also true for the
running in the Yukawa sector from the GUT scale where those
relations hold to the scale relevant for the Yukawa couplings
entering nucleon decay. On the other hand, maximal suppression
corresponds to a case with particular relation between unitary
matrices responsible for bi-unitary transformations in the Yukawa
sector~\cite{Ilja3} that define physical basis for quarks and
leptons.) In both cases we use $\alpha^{-1}_{GUT}=35$ and the best
limit on partial lifetime which is established for $p \rightarrow
\pi^0 e^+$ decay channel ($\tau>5.0 \times 10^{33}$\,years). This
gives conservative bound for the suppressed case since it is
always possible to rotate away proton decay contributions for
individual channels~\cite{Ilja3}.

The uncertainty in extracting the limits on $M_V$ from
experimental data is easy to understand. Namely, even though the
nucleon lifetime is proportional to $M_V^4$, which would make
extraction rather accurate and precise, the lifetime is also
proportional to the fourth power of a term which is basically a
sum of entries of unitary matrices which are \textit{a priori}\/
unknown unless the Yukawa sector of the GUT theory is specified
and which, in magnitude (see Eqs.~\eqref{d=6coefficients}), can
basically vary from $V_{ub}$ to $1$~\cite{Nandi,Ilja3}. (For full
discussion see also Appendix~\ref{second}.) If we now adopt the
$M_{GUT}\equiv M_V$ assumption and use Eq.~\eqref{condition2} the
above limits translate into $B_{12}<7.0$ ($B_{12}<6.1$) for the
suppressed (unsuppressed) case. So, all $SU(5)$ GUTs with
$B_{12}>7.0$ are excluded by the usual gauge $d=6$ contributions
to proton decay. The theories with $6.1 <B_{12}<7.0$ require
``special" structure of the Yukawa sector; the closer the $B_{12}$
to the upper limit is the more ``special" structure is needed.
Finally, any $SU(5)$ GUT with $B_{12}<6.1$ has not yet been probed
by proton decay experiments. (Again, this is all based on the
one-loop analysis. Any more accurate and precise statement must be
based on the two-loop treatment with a proper inclusion of the
threshold corrections.)

In order to avoid problems with proton decay without requiring too
much conspiracy in the Yukawa sector we pursue the solutions where
the superheavy gauge bosons are as heavy as possible. So, how
heavy can they be given the particle content of the $SU(5)$
scenario with an extra $\bm{15}$ of Higgs? In order to answer that
we first naively set masses of $\Sigma_3$, $\Phi_a$ and $\Phi_b$
to $M_Z$. This in turn yields the lowest possible value of
$B_{12}$ to be $6.4$ ($6.33$) for $n=1$ ($n=2$) which translates
via Eq.~\eqref{condition2} into $M_{GUT}=3.2 \times 10^{14}$\,GeV
($M_{GUT}=4.4 \times 10^{14}$\,GeV). (We include $n=2$ case in our
considerations since it might be relevant in addressing Baryon
asymmetry of the Universe.) In this naive estimate $M_V$ is either
equal to or slightly above $M_{GUT}$. From the previous discussion
on the $M_V$ limits we see that there is a need for small
suppression in Yukawa sector in order to satisfy experimental
limits on proton decay. This suppression, as we show later,
amounts to a $1/5$ of the unsuppressed case. When compared to the
available suppression ($\sim 1/V_{ub}$) it comes out to be around
$2 \%$, which can easily be accomplished.

Note, however, that in order to have exact unification crucial
thing is to satisfy Eq.~\eqref{condition1}. Thus, it is better to
ask for which mass spectrum that satisfies Eq.~\eqref{condition1}
we obtain the highest possible value for $M_V$ or equivalently the
smallest possible value for $B_{12}$. As it turns out the answer
to this question is unique within our scenario. To show that we
first assume that the relevant degrees of freedom that improve the
running, i.e., $\Sigma_3$, $\Phi_a$ and $\Phi_b$, contribute in
pairs, e.g., a degenerate pair $(\Sigma_3,\Phi_a)$ is light and
$\Phi_b$ is at $M_{GUT}$, and treat only the $n=2$ case. With
those constraints we generate three possible combinations which
yield results summarized in Table~\ref{tab:table3}. (We address
both the $n=1$ and $n=2$ case at the two-loop level in
Appendix~\ref{third}.)
\begin{table}[h]
\caption{\label{tab:table3} $\Delta B_{23}$ and $\Delta B_{12}$
corrections due to the degenerate pairs of fields and the
associated scales for the $n=2$ case.}
\begin{ruledtabular}
\begin{tabular}{lccc}
     &$(\Sigma_3,\Phi_a)$&$(\Sigma_3,\Phi_b)$ & $(\Phi_a,\Phi_b)$\\
\hline $\Delta B_{23}$& $\frac{6}{6} r+\frac{2}{6}$&$\frac{3}{6}
r+\frac{2}{6}$&$\frac{5}{6} r+\frac{2}{6}$\\
$\Delta B_{12}$&$-\frac{6}{15} r -\frac{2}{15}$&$-\frac{12}{15} r
-\frac{2}{15}$ &$-\frac{8}{15} r -\frac{2}{15}$\\
$M_{GUT}$&$5 \times 10^{13}$\,GeV& ---&$9 \times 10^{13}$\,GeV\\
$M_{r}$&$1$\,TeV& --- &$200$\,GeV\\
\end{tabular}
\end{ruledtabular}
\end{table}

\begin{itemize}
    \item The $(\Sigma_3,\Phi_a)$ case with $n=2$ exactly mimics the $n=8$
case in terms of quantum numbers. (Recall that it takes at least
eight light Higgs doublets on top of the Higssless SM content to
unify the couplings. Associated corrections to the Higssless SM
coefficients are $\Delta B_{23}=\frac{6}{6} r+\frac{2}{6}$ and
$\Delta B_{12}=-\frac{6}{15} r-\frac{2}{15}$, where $r$, as
defined in Eq.~\eqref{r}, is very close to one and we take two of
the doublets to be at $M_Z$.) The unification scale is rather low
and very close to the experimentally set limit for maximally
suppressed case. Lightness of $\Phi_a$ goes against the idea
behind the type II see-saw if one assumes that the parameter $c_3$
(see Eqs.~\eqref{potential} and~\eqref{seesaw}) is at the GUT
scale, but at this point the scenario is \textit{not ruled out}\/
experimentally.
    \item The $(\Phi_a,\Phi_b)$ case has a slightly higher unification scale
than the $(\Sigma_3,\Phi_a)$ case. This time both $\Phi_a$ and
$\Phi_b$ have mass in phenomenologically interesting region.
Lightness of $\Phi_a$ again requires large suppression in the
Yukawa sector for neutrinos to generate correct mass scale via
type II see-saw. However, such a suppression would be beneficial
in suppressing novel contributions to proton decay due to the
mixing between $\Phi_b$ and $\Psi_T$.
    \item The $(\Sigma_3,\Phi_b)$ case is the most promising. Even though it
fails to unify at the one-loop level its correction to $B_{12}$ is
the largest of all three cases. As such, it represents the best
possible candidate to maximize $M_V$. Moreover, the $\Phi_a$
contribution to the running to produce unification for light
$\Sigma_3$ and $\Phi_b$ is small which implies that its mass could
be in the range that is optimal for the type II see-saw for the
most ``natural" value of $c_3$ coefficient. Again, the
$(\Sigma_3,\Phi_b)$ case not only maximizes $M_V$ but also places
$M_{\Phi_a}$ at the right scale to explain neutrino masses.
\end{itemize}

The three special cases discussed above all demonstrate that large
$M_V$ scale prefers $\Phi_b$ light regardless of the relevant
scale of other particles since it is $\Phi_b$ coefficients that
decrease $B_{12}$ the most. These conclusions persist in more
detailed one- and two-loop studies.

Is there a way to tell between the three limiting cases we just
discussed? The $(\Sigma_3,\Phi_a)$ case can be tested and excluded
by slight improvement in the nucleon lifetime data; other low
energy signatures depend on how light $\Phi_a$ is. One could also
test and distinguish between the $(\Sigma_3,\Phi_b)$ and
$(\Phi_a,\Phi_b)$ cases since both favor light $\Phi_b$
leptoquarks that can be detected by LHC. If and when these are
detected the two cases could be distinguished by the scalar
leptoquark contributions to the rare processes. In the
$(\Phi_a,\Phi_b)$ case the suppression in the neutrino Yukawa
sector would selectively erase some of these contribution while in
the $(\Sigma_3,\Phi_b)$ case all these contributions would be
sizable. Expected improvements in the table-top experiments would
then be sufficient to tell the two.

\begin{figure}[h]
\begin{center}
\includegraphics[width=4.5in]{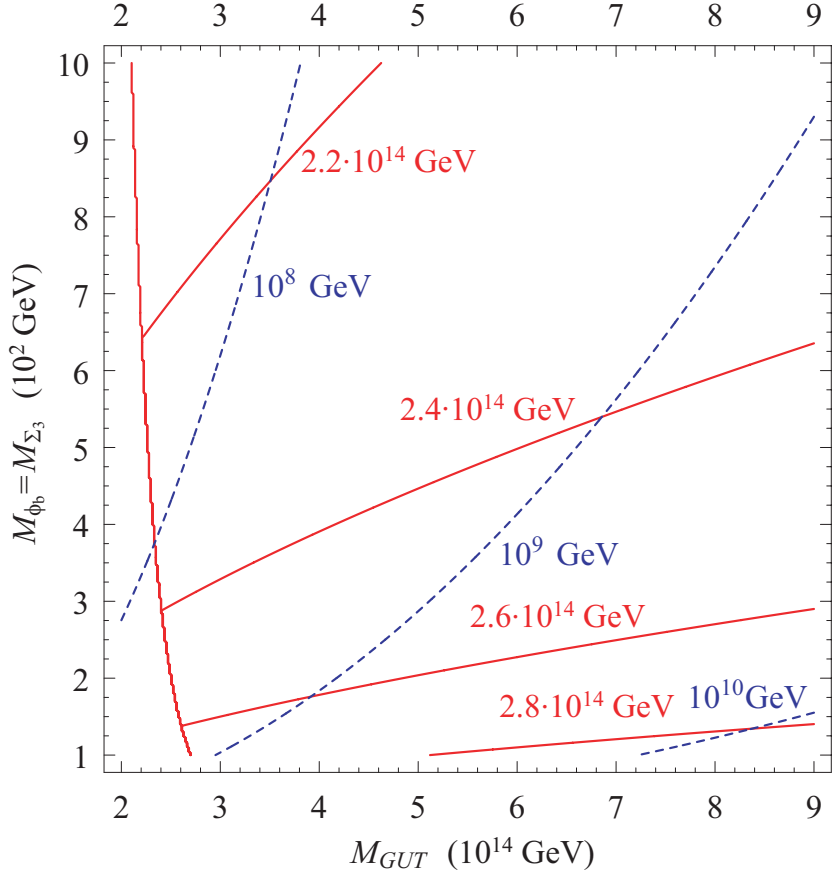}
\end{center}
\caption{\label{figure:1} Plot of lines of constant value of $M_V$
(solid lines) and $M_{\Phi_a}$ (dashed lines) in the
$M_{GUT}$--$(M_{\Sigma_3}=M_{\Phi_b})$ plane. The area to the left
of a steep solid line denotes the region where $M_V$ does not
contribute to the running, i.e., $M_V \geq M_{GUT}$. To generate
the plot we consider exact one-loop unification and use central
values for the gauge couplings as given in the text. This is a
scenario with one light Higgs doublet ($n=1$).}
\end{figure}

\begin{figure}[h]
\begin{center}
\includegraphics[width=4.5in]{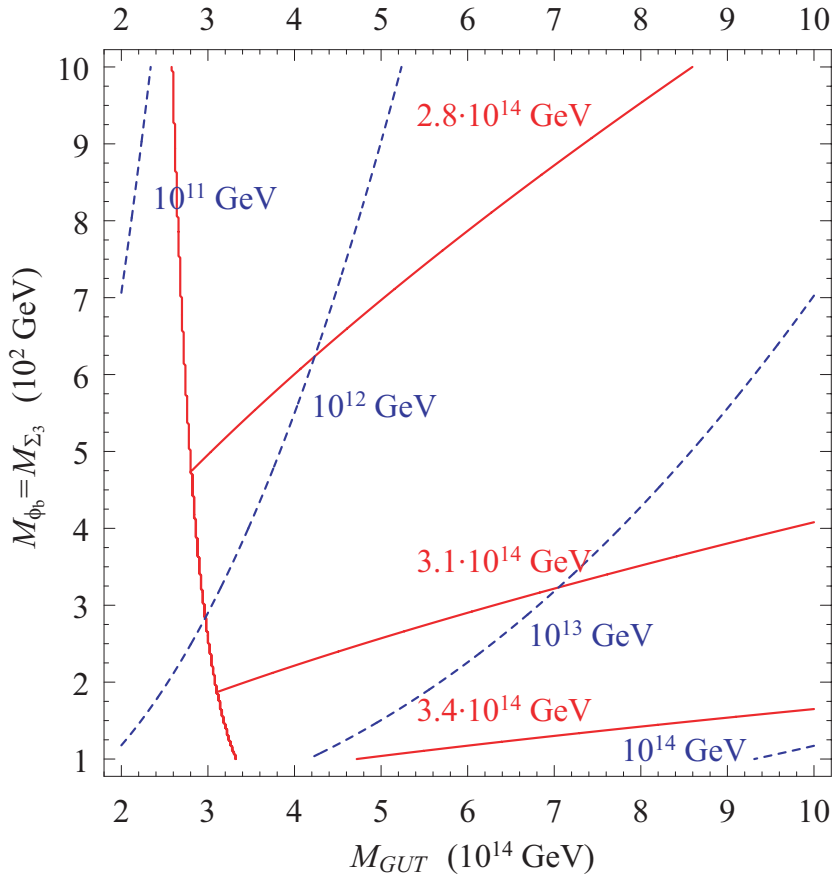}
\end{center}
\caption{\label{figure:2} Plot of lines of constant value of $M_V$
(solid lines) and $M_{\Phi_a}$ (dashed lines) in the
$M_{GUT}$--$(M_{\Sigma_3}=M_{\Phi_b})$ plane. The area to the left
of a steep solid line denotes the region where $M_V$ does not
contribute to the running, i.e., $M_V \geq M_{GUT}$. This is the
$n=2$ scenario.}
\end{figure}

\begin{figure}[h]
\begin{center}
\includegraphics[width=4.5in]{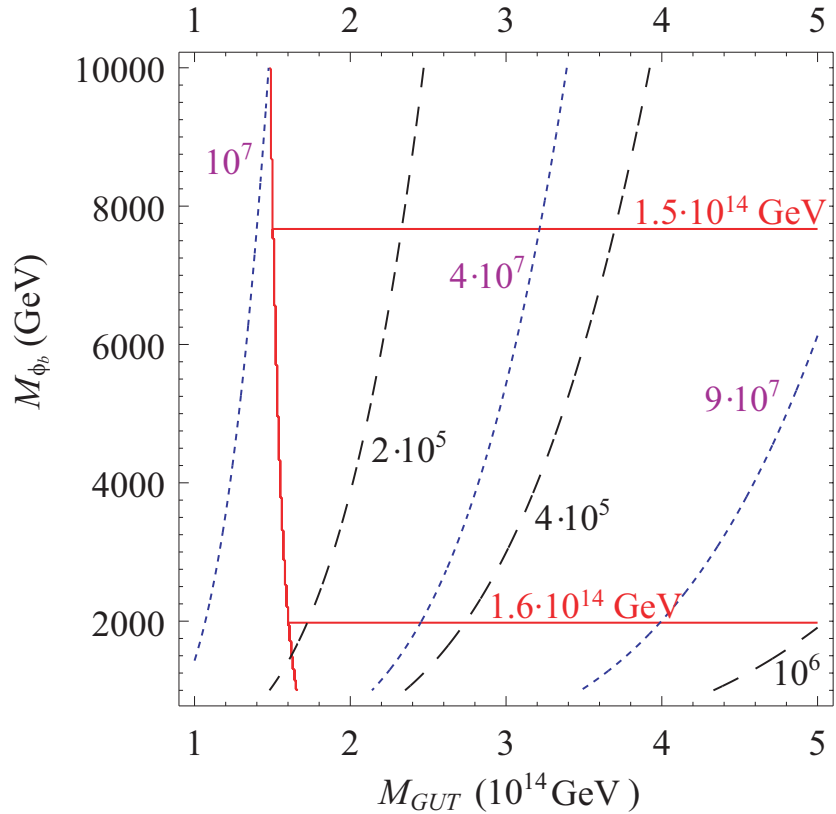}
\end{center}
\caption{\label{figure:3} Plot of lines of constant value of $M_V$
(solid lines) and $M_{\Phi_a}=M_{\Sigma_3}$ in GeV units (dashed
lines) in the $M_{GUT}$--$(M_{\Phi_b})$ plane. The scenario with
one (two) Higgs doublet(s) corresponds to long (short) dashed
lines.}
\end{figure}

Since we have $M_{GUT}$ and masses of $V$, $\Sigma_3$, $\Phi_a$
and $\Phi_b$ as free parameters and only two
equations---Eqs.~\eqref{condition1} and~\eqref{condition2}---we
present four special cases based on certain simplifying
assumptions in Figs.~\ref{figure:1}, \ref{figure:2}
and~\ref{figure:3} and discuss each case in turn. All examples we
present generate consistent unification in agreement with low
energy data. (Note that the change in the parameters also affects
the value of $\alpha^{-1}_{GUT}$. We do not present that change
explicitly, which, for the range of values we use, vary from $36$
to $40$. In our plots we also allow $M_V$ to be at most factor of
three or four lower than the GUT scale. Once we switch to two-loop
analysis with threshold corrections accounted for we appropriately
set $M_V=M_{GUT}$.)

In Fig.~\ref{figure:1} we present the $n=1$ case when the pair
($\Sigma_3$,$\Phi_b$) is taken to be degenerate with the mass
close to electroweak scale ($\sim O(10^{2})$\,GeV). The parameter 
$c_3$ has to be between $10^{3}$ and $10^{7}$ GeV to 
explain the neutrino mass through the type II see-saw if the Yukawa
coupling for neutrinos is of order one. On the other hand, the gauge boson 
mass varies only slightly for a given range of $M_{\Phi_b}$ and
$M_{\Sigma_3}$ around $2.5 \times 10^{14}$\,GeV. Clearly,
unification itself allows $M_{\Phi_b}$ and $M_{\Sigma_3}$ to be
much heavier then 1\,TeV on account of decrease of $M_{\Phi_a}$
but in that case $M_V$ would be getting lighter. This, on the
other hand would require additional conspiracy in the Yukawa
sector in order to sufficiently suppress proton decay to avoid the
experimental limit.

The two light Higgs doublet case is presented in
Fig.~\ref{figure:2}. This case is well motivated on the
Baryogenesis grounds. Namely, the interaction of the $\bm{15}$ of
Higgs explicitly breaks $B-L$ symmetry (see Appendix~\ref{first}).
This opens a door for possible explanation of the Baryon asymmetry
in the Universe within our framework. However, since the
successful generation of Baryon asymmetry requires at least two
Higgses in the fundamental representation we study a consistent
unification picture for that case. (See references for the
Baryogenesis mechanisms in the context of $SU(5)$ model with two
Higgses in the fundamental representation
\cite{Barr,Nanopoulos,Yildiz,Fukugita}). The $n=2$ case has higher
scale of superheavy gauge bosons compared to the $n=1$ case (this
does not hold at the two-loop level though) and the mass of the
field $\Phi_a$ is in the region relevant for the type II see-saw.
Hence, if the mass of scalar leptoquarks is in phenomenologically
interesting region ($\sim O(10^2)$\,GeV) we can explain neutrino
masses naturally. Again, as in the $n=1$ case, the gauge boson
mass varies very slightly, this time around $3 \times
10^{14}$\,GeV.
Given the last two examples we can again conclude that the exact
unification in this minimal realistic scenario points towards
light scalar leptoquarks.

In order to understand better these results we show two more
examples in Fig.~\ref{figure:3}. This time we set
$\Sigma_3=\Phi_a$ for simplicity and present both $n=1$ and $n=2$
cases. This scenario is disfavored by the fact that $M_V$ tends to
be ``small" but cannot be excluded on experimental grounds. It is
evident from the plot that $M_V$ does not depend on a number of
light Higgs doublets and $M_{GUT}$. The reason for that is very
simple and is valid only at the one-loop level. Namely, the ratio
$B_{23}/B_{12}$ is the same for $\Psi_D$ coefficients ($=-5/2$) as
for the sum of corresponding $\Sigma_3$ and $\Phi_a$ coefficients
provided these are degenerate. Thus, any change in the number of
light doublets in Eq.~\eqref{condition1} is simply compensated by
the change in degenerate mass of $\Sigma_3$ and $\Phi_a$ fields
for a fixed value of $\Phi_b$ mass. This trend can be clearly seen
in Fig.~\ref{figure:3}. The mass of $\Phi_a$ is generally rather
low to generate neutrino mass of correct magnitude unless Yukawa
couplings of neutrinos are extremely small.

In all our examples, $\Sigma_3$ is allowed to be much lighter than
$\Sigma_8$. But, this seems in conflict with the tree-level
analysis of the $\Sigma$ potential that is invariant under $\Sigma
\rightarrow -\Sigma$ transformation which yields a well known
relation $M_{\Sigma_3}= 4 M_{\Sigma_8}$~\cite{Buras}. This
apparent mismatch has a simple remedy.

\begin{figure}[h]
\begin{center}
\includegraphics[width=4.5in]{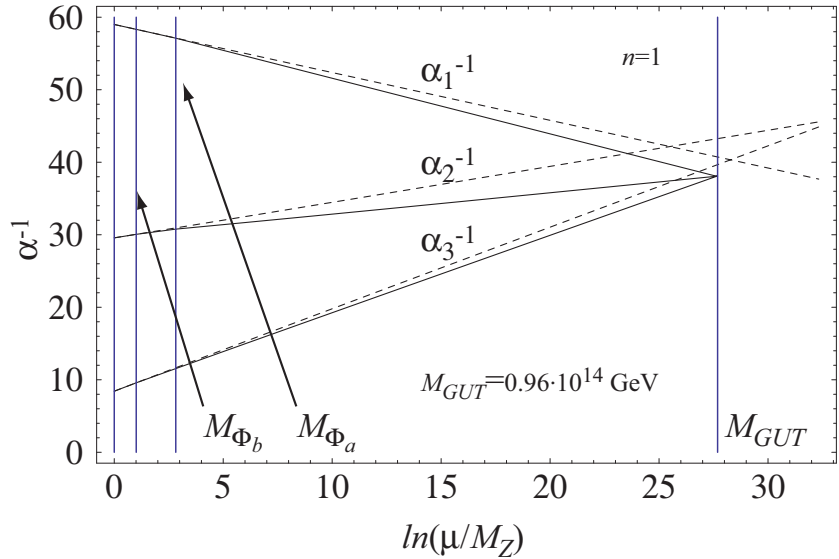}
\end{center}
\caption{\label{figure:4} Unification of the gauge couplings at
the two-loop level for central values of low-energy
observables~\cite{PDG2004}. The SM case with $n=1$ is presented by
dashed lines. Solid lines correspond to the $n=1$ scenario with
$\Phi_b$ and $\Phi_a$ below the GUT scale. Vertical lines mark the
relevant scales: $M_Z$, $M_{\Phi_b}=250$\,GeV,
$M_{\Phi_a}=1.54$\,TeV and $M_{GUT}=0.96 \times 10^{14}$\,GeV.}
\end{figure}

In order to generate sufficiently large corrections to charged
fermion masses via higher-dimensional operators in the Yukawa
sector we need terms linear in $\Sigma/M_{Pl}$. If that is the
case it is no longer possible to require that the Lagrangian is
invariant under transformation $\Sigma \rightarrow -\Sigma$. It is
then necessary to include a cubic term into the $\Sigma$ potential
besides the usual quadratic and quartic ones. But, the potential
with the cubic term ($Tr \Sigma^3$) violates the validity of
$M_{\Sigma_3}= 4 M_{\Sigma_8}$ relation~\cite{Li,Guth:1,Guth:2}
and allows a possibility where $\Sigma_3$ is light while
$\Sigma_8$ is superheavy. We analyze this situation in
Appendix~\ref{first} in some detail. Note that we do not require
nor insist on the lightness of $\Sigma_3$ though. In
Appendix~\ref{third} we present the two-loop analysis of the
scenario where $\Phi_a$ is relatively light ($\sim 10^{7}$\,GeV)
and $\Sigma_3$ is at the GUT scale. Our intention is solely to
demonstrate that there are more possibilities available unless
additional assumptions, such as $\Sigma \rightarrow - \Sigma$
transformation, are imposed on the $SU(5)$ theory. Note, however,
that the maximization of $M_V$ \textit{always}\/ requires $\Phi_b$
to be very light ($\sim 10^{2}$\,GeV). In Fig.~\ref{figure:4} we show
an example where it is possible to achieve unification at
the two-loop level (See details in Appendix~\ref{third}) for $n=1$, 
$M_{\Phi_b}=250$\,GeV, $M_{\Phi_a}=1.54$\,TeV and the
field $\Sigma_3$ is at the GUT scale. 
  
What about the possible mass spectrum of $\Phi_a$, $\Phi_b$ and
$\Phi_c$? The relevant potential is in Appendix~\ref{first}.
Clearly, there are more parameters than mass eigenvalues. The
tree-level analysis revels that it is possible to obtain any
possible arrangement including, for example, $\Phi_b \ << \ \Phi_a
\ < \ \Phi_c$. This sort of split is quite similar to the split
behind the well-known Doublet-Triplet problem.

Our framework yields rather low mass for vector leptoquarks that
varies within very narrow range around $3 \times 10^{14}$\,GeV for
a most plausible scenarios. This makes the framework testable
through nucleon decay measurements. (More precisely, large portion
of the parameter space of the setup has already been excluded by
existing measurements on nucleon lifetime.) It is easy to
understand this generic and robust prediction. We have seen that
$M_{GUT}$ can be at most $3.2 \times 10^{14}$\,GeV ($4.4 \times
10^{14}$\,GeV) if we exclude the $V$ contribution from the running
for the $n=1$ ($n=2$) case. If we now start lowering $M_V$ below
$M_{GUT}$ we lower $B_{12}$ as well. This, on the other hand,
starts to increase $M_{GUT}$ but still keeps $M_V$ at almost the
same value. Basically, the ``decoupling" of $M_V$ and $M_{GUT}$
takes place, where the mass of $X$ and $Y$ gauge bosons remains in
vicinity of the value before the ``decoupling" while $M_{GUT}$
rapidly approaches the Planck scale. Another way to say this is
that the $B_{ij}$ coefficients of superheavy gauge fields are very
large compared to all other relevant coefficients (see
Table~\ref{tab:table2}). Thus, any small change in $M_V$
corresponds to a large change in other running parameters.

Let us finally investigate possible experimental signatures coming
from our consistent minimal realistic $SU(5)$ model.
\section{Experimental Signatures}
\label{IV} Our framework has potential to be tested through the
detection---direct or indirect---of light leptoquarks and/or
observation of proton decay. Let us investigate each of these
tests in turn.

\begin{itemize}

\item \textbf{Light leptoquarks}

To get consistent unification in agreement with low energy data,
neutrino mass and proton decay in our minimal framework we
generate very light leptoquarks $\Phi_b$. The lighter the $\Phi_b$
is the heavier the $V$ becomes. Thus, in the most optimistic
scenario $M_{\Phi_b}$ is close to the present experimental limit
$\sim O(10^{2})$\,GeV. In what follows we specify all relevant
properties of $\Phi_b$ and existing constraints on its couplings
and mass.

The $\bar{\bm{5}}_a^T C h_{ab} \Phi \bar{\bm{5}}_b$ coupling
yields the following interactions:
\begin{equation}
\label{leptoquark} {d^C}^T C h \Phi_b l = {d^C}^T C h ( \phi_b^1 e
- \phi_b^2 \nu ),
\end{equation}
where the leptoquarks $\phi_b^1$ and $\phi_b^2$ have electric
charges $2/3$ and $-1/3$, respectively, and symmetric matrix $h$
coincides with the Yukawa coupling matrix of Majorana neutrinos
($\equiv Y_\nu$) if we neglect the Planck suppressed operators.
(See the last line in Eq.~\eqref{yukawa}.) The above leptoquark
interactions in the physical basis read as:
\begin{equation}
{d^C}^T \ C \ (D_C^T E^*) K^* (V_{PMNS}^M)^*
Y_{\nu}^{\textrm{diag}}(V_{PMNS}^M)^{\dagger} K^* \ \phi_b^1 \ e,
\end{equation}
\begin{equation}
{d^C}^T \ C \ (D_C^T E^*) K^* (V_{PMNS}^M)^*
Y_{\nu}^{\textrm{diag}} \ \phi_b^2 \ \nu,
\end{equation}
where $D_C$ and $E$ are the matrices which act on $d^C$ quarks and
$e$, respectively, to bring them into physical basis. (See
Appendix~\ref{second} for exact convention.) $K$ is a matrix
containing three CP violating phases, and $V_{PMNS}^M$ is the
leptonic mixing in the Majorana case. (In the GG $SU(5)$ where
$Y_E = Y_D^T$ one has $D_C = E$. However, that is not the case in
a realistic model for fermion masses.)

There are many studies about the contributions of scalar and
vector leptoquarks in different
processes~\cite{Buchmuller,Davidson}. For a model independent
constraints on leptoquarks from rare processes see for
example~\cite{Davidson,Herz}. The most stringent bound on the
scalar leptoquark coupling to matter comes from the limits on
$\mu$--$e$ conversion on nuclei~\cite{Dohmen}. The bound we
present should be multiplied by
$(M_{\Phi_b}/100\,\textrm{GeV})^2$. In our case it reads $(D_C^{T}
h E)_{11}(E^\dagger h^\dagger D_C^{*})_{21}<10^{-6}$. The bounds
for all other elements of $(D_C^{T} h E )_{ij}(E^\dagger h^\dagger
D_C^{*})_{kl}$ and $(D_C^{T} h N)_{ij}(N^* h^\dagger
D_C^{*})_{kl}$ are weaker.

The currents bounds on leptoquarks production are set by Tevatron,
LEP and HERA~\cite{bounds}. Tevatron experiments have set limits
on scalars leptoquarks with couplings to $eq$ of $M_{LQ} >
242$\,GeV. The LEP and HERA experiments have set limits which are
model dependent. The search for these novel particles will be
continued soon at the CERN LHC. Preliminary studies by the LHC
experiments ATLAS~\cite{ATLAS} and CMS~\cite{CMS} indicate that
clear signals can be established for masses up to about $M_{LQ}
\approx 1.3 $\,TeV. For several studies about production of scalar
leptoquarks at the LHC, see Ref.~\cite{production}. Thus, it could
be possible to test our scenario at the next generation of
colliders, particularly in the Large Hadron Collider (LHC) at
CERN, through the production of light leptoquarks. Therefore even
without the proton decay experiments we could have tests of this
non-supersymmetric GUT scenario.

\item \textbf{Proton decay}

Proton decay is the most generic prediction coming from matter
unification; therefore, it is the most promising test for any
grand unified theory. (For new experimental bounds
see~\cite{PDG2004,newbounds}.) In our minimal and consistent
scenario the relevant scale for gauge bosons is around $3 \times
10^{14}$\,GeV regardless of how high the GUT scale goes in order
to get consistent unification in agreement with low energy data.
Careful study within the two-loop context with the inclusion of
threshold effects revels that the highest possible value of $M_V$
in the $n=2$ ($n=1$) case is $3.19 \times 10^{14}$\,GeV ($3.28
\times 10^{14}$\,GeV) for central values of coupling
constants~\cite{PDG2004} while the $1\,\sigma$ departure allows
for the maximum value of $3.35 \times 10^{14}$\,GeV in the $n=2$
case, for example.

There are several contributions to the decay of the proton in our
minimal scenario. We have the usual Higgs and gauge $d=6$
operators but there are also new contributions due to the mixing
between $\Psi_T$ and $\Phi_b$, with $\Phi_b$ being extremely light
in our case. These contributions are very important. Using the
relevant triplet interactions:
\begin{equation}
q_a \ C \ \underline{A}^{ab} \ q_b \ \Psi_T \ + \ q_a \ C \ \underline{C}^{ab}
\ l_b \ \Psi_T^* \ + \ u^C_a \ C \ \underline{D}^{ab} \ d_b^C \
\Psi_T^* \ + \ u^C_a \ C \ \underline{B}^{ab} \ e_b^C \ \Psi_T,
\end{equation}
(for the expressions of $\underline{A}$, $\underline{B}$,
$\underline{C}$, and $\underline{D}$ matrices see for
example~\cite{Bajc2}) and the interaction term $\bar{\bm{5}}_a^T C
h_{ab} \Phi \bar{\bm{5}}_b$, it is easy to write down the
contributions for the $B-L$ non-conserving decays $p \to (K^+,
\pi^+, \rho^+) \nu_i$, and $n \to ( \pi^0, \rho^0, \eta^0, w^0,
K^0) \nu_i$. We present the relevant diagram in
Fig.~\ref{figure:5}~\cite{Mohapatrabook}.
\begin{figure}[h]
\begin{center}
\includegraphics[width=3in]{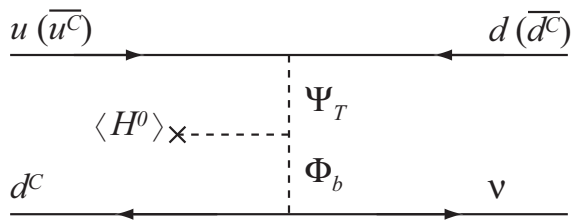}
\end{center}
\caption{\label{figure:5} Contributions to the decay of the proton
induced by the $\bm{15}$ of Higgs.}
\end{figure}

Notice that in this scenario we have the usual $B-L$ conserving
decays, i.e., the decays into a meson and antileptons, and the
$B-L$ non-conserving decays mentioned above. Since $\Phi_b$ has to
be light, the $B-L$ violating decays are very important. The rate
for the decays into neutrinos is given by:
\begin{eqnarray}
\Gamma( p \to K^+ {\nu}) & = & \sum_{i=1}^3 \Gamma( p \to K^+
{\nu}_i) = \frac{(m_p^2 - m_K^2)^2}{32 \pi m_P^3 f_{\pi}^2 } A_L^2
\frac{c_3^2 <H^0>^2}{M_{\Psi_T}^4 M_{\Phi_b}^4} \times \nonumber \\
& &
 | \tilde{\beta} \ C(\nu_i, s, d^C) \ + \ \tilde{\alpha} \ C(\nu_i, s^C,
  d^C)|^2 \ \frac{4 m_p^2 D^2}{9 m_B^2} \nonumber \\ &+& | \tilde{\beta} \ C(\nu_i, d,
  s^C) \ + \ \tilde{\alpha} \ C(\nu_i, d^C, s^C)|^2  \ [1 + \frac{m_p (D + 3 F)}{3 m_B}]^2
\end{eqnarray}
where $C(\nu_i, d_{\alpha}, d_{\beta}^C) = (U^T \ ( \underline{A}
+ \underline{A}^T ) D)_{1 \alpha}$ and $C(\nu_i, d_{\alpha}^C,
d_{\beta}^C)= (D_C^{\dagger} \underline{D}^{\dagger}
U_C^*)_{\alpha 1} $. (See Appendix B for notation.)

As you can appreciate from the above expressions, the predictions
coming from these contributions are quite model dependent.
Using $m_p=938.3$\,MeV, $D=0.81$, $F=0.44$, $m_B=1150$\,MeV,
$f_{\pi}=139$\,MeV, $A_L=1.43$, and
$\tilde{\alpha}=\tilde{\beta}=\alpha=0.003$\,GeV$^3$ we get:
\begin{eqnarray}
1.95 \times  10^{-64} \ \text{GeV}^{-6} \ \frac{M_{\Psi_T}^4
  M_{\Phi_b}^4 }{ c_3^2} \ > \ [ \ 0.19 \ | C(\nu_i, s, d^C) \ + \ C(\nu_i, s^C,
  d^C)|^2  \ + \nonumber \\
 + \  2.49 \ | C(\nu_i, d, s^C) \ + \ C(\nu_i, d^C, s^C)|^2 \ ]
\end{eqnarray}
Let us see an example, using the values $M_{\Psi_T} = c_3 = 10^{14}$ GeV and
$M_{\Phi_b}=10^{3}$ GeV, the left-hand side of the above equation
is equal to $1.95 \times 10^{-24}$; therefore, the sum of the $C$
coefficients has to be basically $10^{-12}$. In the case that 
coefficient $c_3$ is smaller ($\sim 10^{6}$\,GeV), a possibility 
that is not excluded, the sum of $C$
coefficients would be around $10^{-4}$, which is their ``natural"
value. Moreover, the scenario would then prefer $\Phi_a$ at the
same scale ($\sim 10^{6}$\,GeV) if $Y_\nu$ is taken to be
proportional to $Y_e$.
Also we can suppress the relevant contributions 
in different ways. For example, we could 
choose $\underline{A}_{ij} = - \underline{A}_{ji}$ and 
$\underline{D}_{ij}=0$ except for $i = j = 3$, or set to zero 
these coefficients in specific models for fermion masses. 

In any case, if the gauge $d=6$ contributions are the dominant
ones for proton decay, we can get the following bounds for the
proton lifetime (see Appendix~\ref{second}) allowing the full
freedom in the Yukawa sector:
\begin{equation}
1 \times 10^{31}\,\text{years} < \tau_p < 2 \times
10^{38}\,\text{years}.
\end{equation}
Here we use $\alpha_{GUT}^{-1}=39$ and $M_{GUT}=3.2 \times
10^{14}$\,GeV. (See Fig.~\ref{figure:6} for example.) Having in
mind the experimental limit of $5.0 \times
10^{33}$\,years~\cite{PDG2004} we see that a significant portion
of available parameter space has already been excluded. We hope
that in the next generation of proton decay experiments this
scenario would be constrained even further.
\end{itemize}

How does this scenario compare to other possible extensions of the
GG model? We mention only few listing them by increasing order in
particle number.
\begin{itemize}

\item The most obvious extension is to add one more fundamental
representation in the Higgs sector to the $n=2$ scenario we
analyze the most at the one-loop level. This addition would not
raise but actually lower $M_V$ since the scalar leptoquarks which
influence $B_{12}$ the most would get slightly heavier than in the
$n=2$ case. In certain way, this actually makes the $n=1$ scenario
with the $\bm{15}$ of Higgs very unique. It is the minimal
extension of the GG model with the highest available scale for
$M_V$. We focus our attention on the $n=2$ case on the grounds of
Baryogenesis. In the same manner, the $n=3$ scenario would be well
motivated by the possibility of addressing the issue of the SM
model CP violation~\cite{Branco0} (see also~\cite{Branco} and
references therein).

\item Very interesting possibility would be the $n=1$ case with
two $\bm{10}$s of Higgs. Such a scenario could have a very high GUT
scale and still very promising phenomenological consequences due
to light leptoquarks. For example, successful two-loop unification
with light $\Phi_b$s (there are two now) at 250\,GeV requires
$M_{\Sigma_3}=2.1 \times 10^{11}$\,GeV and the GUT scale at $1.0
\times 10^{15}$\,GeV for central values of $\alpha_i$s at $M_Z$.
This model would also require at least right-handed neutrinos and
non-renormalizable operators to be completely realistic.

\item Another possibility would be the $n=1$ case with one
$\bm{10}$ and one $\bm{15}$ of Higgs. Such a scenario would have a
same maximal value for the GUT scale as the two $\bm{10}$ case
above. To be completely realistic it would require
non-renormalizable operators.

\item The next scenario is the one proposed by Murayama and
Yanagida~\cite{MY} (MY). They shown that addition of two
$\bm{10}$s of Higgs to the GG model and with $n=2$ it is possible
to achieve unification for extremely light scalar leptoquarks in
the $\bm{10}$s. This allows MY to forward a ``desert" hypothesis,
within which the particles are either light ($\sim O(10^2)$\,GeV)
or heavy ($\sim O(10^{14})$\,GeV). (Note that the scalar
leptoquarks with exactly the same quantum numbers as $\Phi_b$
($\subset \bm{15}$) reside in the $\bm{10}$ as well. To see that
one can use $\bm{5} \otimes \bm{5}=\bm{10}\oplus\bm{15}$. Note
that the couplings of $\Phi_b$ in the $\bm{10}$ ($\bm{15}$) to the
$\bm{5}$s are antisymmetric (symmetric).) Their model is ruled out
by direct searches for scalar leptoquarks due to the extreme
lightness of $\Phi_b$s. However, if one allows for splitting
between $\Sigma_3$ and $\Sigma_8$ one can raise unification scale
sufficiently to avoid proton decay bounds and resurrect their
model although one would have to abandon the ``desert" hypothesis
MY forwarded. The additional contribution to the Higgsless SM
coefficient in this case is basically $\Delta
B_{12}=-\frac{14}{15} r_{\Phi_b}-\frac{5}{15}
r_{\Sigma_3}-\frac{2}{15}$ which is much more than any of the
cases considered in Table~\ref{tab:table3}. For example, for
$M_{\Phi_b}=300$\,GeV we find, at the one-loop level,
$M_{\Sigma_3}=10^{10}$\,GeV and $M_{GUT}=9 \times 10^{14}$\,GeV.
Note that the presence of the $\bm{10}$ of Higgs yields the same
type of coupling as we specify in Eq.~\eqref{leptoquark} except
that $h$, in this case, would be antisymmetric. This, however,
would not \textit{a priori\/} prevent proton decay. If there are
two or more Higgses in the fundamental representation in the model
there exist the proton decay process schematically represented in
Fig.~\ref{figure:4} unless additional symmetry is introduced.
Finally, this model would also require at least right-handed
neutrinos and non-renormalizable operators to be completely
realistic.

\item The Georgi-Jarlskog model~\cite{GJ} with an extra $\bm{45}$
of Higgs to fix fermion masses represents natural extension of the
GG model. However, even though unification takes
place~\cite{BabuMa,Giveon} and the extension has good motivation
the predictivity of the model is lost unless additional
assumptions are introduced. More minimal extension than that would
be, for example, an extra $\bm{24}$ of Higgs to the scenario we
consider.
\end{itemize}
\section{Summary}
\label{conclusions} We have investigated the possibility to get a
consistent unification picture in agreement with low energy data,
neutrino mass and proton decay in the context of the minimal
realistic non-supersymmetric $SU(5)$ scenario. This scenario is
the Georgi-Glashow model extended by an extra $\bm{15}$ of Higgs.
As generic predictions from the running of the gauge couplings we
have that a set of scalar leptoquarks is light, with their mass,
in the most optimistic case, being around
$O(10^{2}$--$10^{3})$\,GeV. This makes possible the tests of this
scenario at the next generation of collider experiments,
particularly in the Large Hadron Collider (LHC) at CERN. In the
``least" optimistic scenario the mass of the scalar leptoquarks
would be around $10^{5}$\,GeV. The proton decay issue has been
studied in detail, showing that it is possible to satisfy all
experimental bounds with very small, at $2\%$ level, suppression
in Yukawa sector. Rather low scale of vector leptoquarks ($\sim 3
\times 10^{14}$\,GeV) already allows for significant exclusion of
available parameter space of our scenario. Further reduction is
expected in near future with new limits on proton decay lifetime.
We have particularly studied the case with two Higgses in the
fundamental representation since in this case it could be possible
to explain the Baryon asymmetry of the Universe. We have also
compared this scenario with other, well motivated, extensions of
the GG model. There are uncertainties related to predictions of
the proposed scenario but, in view of the fact that it truly
represents the minimal realistic extension of the GG model, the
same is even more true of all other extensions.
\begin{acknowledgments}
We would like to thank Stephen M.~Barr, Borut Bajc, M.~A.~Diaz,
Goran Senjanovi\'c and Qaisar Shafi for useful discussions and
comments. I.D.\ would also like to thank Trudy and Arthur on their
support. P.F.P.\ thanks the Max Planck Institute for Physics
(Werner-Heisenberg-Institute) for warm hospitality. I.D.\ thanks
the Bartol Research Institute for hospitality. The work of P.F.P.\
has been supported by CONICYT/FONDECYT under contract
$N^{\underline 0} \ 3050068$.
\end{acknowledgments}
\appendix
\section{Particle Content And Relevant Interactions}
\label{first} In this appendix we define a minimal realistic
non-supersymmetric $SU(5)$ model. By minimal we refer to the
minimal number of physical fields such a model requires. In the GG
model the matter is unified in two representations:
$\bar{\bm{5}}_a=(d^C, l)_a$, and $\bm{10}_a=(u^C, q, e^C)_a$,
where $a=1,2,3$ is a generation index. The Higgs sector comprises
$\mathbf{{5}}_H=\Psi=(\Psi_D, \Psi_T)$ and
$\mathbf{24}_H=\Sigma=(\Sigma_8, \Sigma_3, \Sigma_{(3,2)},
\Sigma_{(\bar{3}, 2)}, \Sigma_{24})$. However, this model does not
achieve unification and fails to correctly accommodate fermion
masses; therefore, it is ruled out. In the introduction we discuss
how it is possible to solve all phenomenological problems of the
GG model introducing a minimal set of fields. Namely, it is
sufficient to introduce an extra representation:
$\mathbf{15}_H=\Phi=(\Phi_a, \Phi_b, \Phi_c)$. The SM
decomposition of the Higgs sector is given by:
\begin{eqnarray}
\bm{24}_H&=& \Sigma =
(\bm{8},\bm{1},0)+(\bm{1},\bm{3},0)+(\bm{3},\bm{2},-5/6)+(\overline{\bm{3}},\bm{2},5/6)+(\bm{1},\bm{1},0),\\
\bm{15}_H&=& \Phi =
(\bm{1},\bm{3},1)+(\bm{3},\bm{2},1/6)+(\bm{6},\bm{1},-2/3),\\
\bm{5}_H&=& \Psi=(\bm{1},\bm{2},1/2)+(\bm{3},\bm{1},-1/3).
\end{eqnarray}
The relevant Yukawa potential, up to order $1/M_{Pl}$, is
\begin{eqnarray}
\label{yukawa} V_\textrm{Yukawa}&=&
\epsilon_{ijklm}\left(\bm{10}_a^{ij}f_{ab}\bm{10}_b^{kl} \Psi^m+
\bm{10}_a^{ij}f_{1ab}\bm{10}_b^{kl}
\frac{\Sigma^m_{~n}}{M_{Pl}}\Psi^n+
\bm{10}_a^{ij}f_{2ab}\bm{10}_b^{kn}\Psi^l \frac{\Sigma^m_{~n}}{M_{Pl}}\right)\nonumber\\
&+&\Psi^*_i \bm{10}_a^{ij}g_{ab}\bar{\bm{5}}_{bj} + \Psi^*_i
\frac{\Sigma^i_{~j}}{M_{Pl}} \bm{10}_a^{jk}g_{1ab}
\bar{\bm{5}}_{bk}
+ \Psi^*_i \bm{10}_a^{ij}g_{2ab}\frac{\Sigma_{~j}^k}{M_{Pl}} \bar{\bm{5}}_{bk} \nonumber\\
& + & \Phi^{ij} \bar{\bm{5}}_{ai} h_{ab} \bar{\bm{5}}_{bj} +
\frac{(\bar{\bm{5}}_{ai} \Psi^i)h_{1ab}(\bar{\bm{5}}_{bj} \Psi^j
)}{M_{Pl}},
\end{eqnarray}
where $i,j,k,l,m$ represent $SU(5)$ indices. Impact of the
non-renormalizable operators on the fermion masses is discussed
in~\cite{Ellis:1}. Notice that we could replace $M_{Pl}$ by a
scale $\Lambda$, where $M_{GUT} \ < \ \Lambda \ < \ M_{Pl}$, if we
do not assume a desert between the GUT scale and the Planck scale.
(See for example reference~\cite{Shafi}.)

The Higgs scalar potential, manifestly invariant under $SU(5)$, is (see for
example~\cite{Ruegg,Buccella}):
\begin{eqnarray}
\label{potential} V_{\textrm{Higgs}}&= &- \frac{\mu_{\Sigma}^2}{2}
\Sigma^i_{~j} \Sigma^j_{~i} +\frac{a_\Sigma}{4} (\Sigma^i_{~j}
\Sigma^j_{~i})^2+\frac{b_\Sigma}{2} \Sigma^i_{~j}
\Sigma^j_{~k}\Sigma^k_{~l} \Sigma^l_{~i}+\frac{c_\Sigma}{3}
\Sigma^i_{~j} \Sigma^j_{~k} \Sigma^k_{~i}
\nonumber\\
& -&\frac{\mu_\Psi^2}{2} \Psi^*_i \Psi^i+ \frac{a_\Psi}{4}
(\Psi^*_i \Psi^i)^2 -  \frac{\mu_\Phi^2}{2} \Phi_{ij}^{*}
\Phi^{ij} + \frac{a_\Phi}{4} (\Phi_{ij}^{*} \Phi^{ij})^2 +
\frac{b_\Phi}{2} \Phi_{ij}^{*} \Phi^{jk} \Phi_{kl}^{*} \Phi^{li}
\nonumber\\
& + & c_1 \Psi_{i}^* \Sigma^i_{~j}  \Psi^j + c_2 \Phi_{ij}^*
\Sigma^j_{~k} \Phi^{ki} +c_3 \Psi^*_{i} \Phi^{ij} \Psi^*_{j} +
c_3^* \Psi^{i}
\Phi^*_{ij} \Psi^{j}\nonumber\\
& + & b_1 \Phi_{ij}^{*} \Phi^{ij} \Sigma^k_{~l} \Sigma^l_{~k} +
b_2 \Psi_i^* \Psi^i
 \Sigma^j_{~k} \Sigma^k_{~j} + b_3 \Psi_i^* \Psi^i
 \Phi_{jk}^{*} \Phi^{jk} +
b_4 \Psi_i^* \Sigma^i_{~j} \Sigma^j_{~k} \Psi^k \nonumber\\
&+& b_5 \Psi^*_{i} \Phi^{ij} \Phi^*_{jk} \Psi^{k}+ b_6 \Phi^{ij}
\Phi^*_{jk} \Sigma^k_{~l} \Sigma^l_{~i}+b_7 \Phi^*_{ij}
\Sigma^j_{~k} \Phi^{kl} \Sigma^i_{~l}.
\end{eqnarray}
We assume no additional global symmetries. It is easy to
generalize this potential to describe the case of two or more
Higgses in the fundamental representation. (We do not include
non-renormalizable terms in the Higgs potential since the split
between $\Sigma_3$ and $\Sigma_8$ masses that we frequently use in
running can already be achieved at the renormalizable level.)

The condition that the symmetry breaking to the SM is a local
minimum of the Higgs potential for $\Sigma$ (the first line in
Eq.~\eqref{potential}) is~\cite{Guth:2}
\begin{equation}
\beta > \left\{
\begin{array}{rc}
\frac{15}{32}(\gamma-\frac{4}{15}), & \gamma>\frac{2}{15} \\
-\frac{1}{16}, & \gamma=\frac{2}{15} \\
-\frac{1}{120 \gamma}, & 0<\gamma<\frac{2}{15} \\
\end{array} \right.
\end{equation}
where dimensionless variables are defined as
\begin{equation*}
\beta=\frac{\mu_{\Sigma}^2 b_\Sigma}{c_\Sigma^2}, \qquad \qquad
\gamma=(\frac{a_\Sigma}{b_\Sigma}+\frac{7}{15}).
\end{equation*}
The vacuum expectation value of $\Sigma$ is $\langle
\Sigma\rangle=\lambda/\sqrt{30} \ \textrm{diag}(2,2,2,-3,-3)$,
where~\cite{Guth:2}
\begin{equation}
\lambda=\frac{c_\Sigma}{b_\Sigma} \left(\frac{\beta}{\gamma}
\right)^{1/2} \left[\left(1+\frac{1}{120 \beta
\gamma}\right)^{1/2}+\frac{1}{(120 \beta
\gamma)^{1/2}}\right]=\frac{c_\Sigma}{b_\Sigma}
\left(\frac{\beta}{\gamma} \right)^{1/2} h(\beta \gamma).
\end{equation}
Finally, the mass of $X$ and $Y$ gauge bosons is given by
\begin{equation}
M_V=\sqrt{\frac{5}{12}} \ g_{GUT} \lambda
\end{equation}
and
\begin{eqnarray*}
M^2_{\Sigma_8}&=& \left[\frac{1}{3}+\frac{5}{\sqrt{30}}
\left(\frac{\gamma}{\beta} \right)^{1/2} \frac{1}{h(\beta \gamma)}
\right] b \lambda^2,\\
M^2_{\Sigma_3}&=& \left[\frac{4}{3}-\frac{5}{\sqrt{30}}
\left(\frac{\gamma}{\beta} \right)^{1/2} \frac{1}{h(\beta \gamma)}
\right] b \lambda^2,\\
M^2_{\Sigma_{24}}&=& \left[1-\frac{1}{1+(1+120 \beta
\gamma)^{1/2}}\right] 2 b \gamma \lambda^2.
\end{eqnarray*}
Here we note the following. In the limit that $c_\Sigma
\rightarrow 0$ we obtain the well know results: $M^2_{\Sigma_3}=4
M^2_{\Sigma_8}$~\cite{Buras}. However, in the limit where $
\lambda \rightarrow \frac{c_\Sigma}{b_\Sigma}\frac{\sqrt{30}}{8}$
we obtain
\begin{eqnarray*}
M^2_{\Sigma_8} &\rightarrow& 5/3 b_\Sigma \lambda^2 \left(\leq
\frac{b_\Sigma}{\pi} \frac{M^2_V}{\alpha_{GUT}}\right),\\
M^2_{\Sigma_3} &\rightarrow& 0,\\
M^2_{\Sigma_{24}} &\rightarrow& \left[\gamma - \frac{2}{5} \right]
2 b \lambda^2.
\end{eqnarray*}
Clearly, it is technically possible to achieve a large split
between $M_{\Sigma_3}$ and $M_{\Sigma_{24}}$ although this is
highly unnatural.

The relevant interactions for the see-saw mechanism are the
following:
\begin{equation}
V_\textrm{see-saw} = - M_{\phi_a}^2 \ \text{Tr} \ \Phi_a^{\dagger}
\Phi_a - Y_{\nu} \ l^T \ C  \ \Phi_a \ l \ + \ c_3 \ \Psi_D^T
\Phi_a^{\dagger} \Psi_D \ + \ \text{h.c.}
\end{equation}
where:
\begin{equation}
\mathbf{\Phi}_a= \left( \begin{array} {cc}
\delta^0 &  - \frac{\delta^+}{\sqrt{2}}\\
- \frac{\delta^+}{\sqrt{2}} & \delta^{++}
\end{array} \right), \qquad
l = \left( \begin{array} {c} \nu \\ e \end{array} \right), \qquad
\Psi_D^T=(H^0 , H^+).
\end{equation}
In this case the neutrino mass is given by:
\begin{equation}
\label{seesaw} M_{\nu} \approx Y_{\nu} \ c_3  \
\frac{<H^0>^2}{M_{\phi_a}^2}
\end{equation}
This is the so-called type II see-saw~\cite{Lazarides,Mohapatra}.
Notice that from this equation in order to satisfy the neutrino
mass experimental constraint ${M_{\phi_a}^2}/{c_3}$ has to be around
$10^{13-14}$\,GeV.

\section{Proton lifetime bounds}
\label{second}
The dominant contribution towards nucleon decay in
non-supersymmetric GUT usually comes from the gauge $d=6$ proton
decay operators. Even though these operators carry certain model
dependence we have recently shown that unlike in the case of $d=5$
operators we can still establish very firm absolute bounds on
their strength that are equally valid for all unifying gauge
groups~\cite{Ilja3}. We investigate the impact these bounds have
on the non-supersymmetric grand unified model building in what
follows.

The upper bound for the total nucleon lifetime~\cite{Ilja3} in
grandunifying theories, in the Majorana neutrino case, reads
\begin{equation}
\tau_p \leq 1.1 \times 10^{41}\,\text{years}
\frac{M^4_V}{\alpha^2_{GUT}},
\end{equation}
where $M_V$---the mass of the superheavy gauge bosons---is given
in units of $10^{16}$\,GeV. We stress that there exist no upper
bound for partial lifetimes since we can always set to zero the decay
rate for a given channel.

In order to fully understand the implications of the experimental
results we now specify the theoretical lower bounds on the nucleon
decay in GUTs. These are applicable in the case of
non-supersymmetric models if the gauge $d=6$ contributions are the
dominant as well as in the case of supersymmetric models where the
$d=4$ and $d=5$ operators are either forbidden or highly
suppressed.

The relevant coefficients for the proton decay amplitudes in the
physical basis of matter fields are~\cite{Pavel}:
\begin{subequations}
\label{d=6coefficients}
\begin{eqnarray}
\label{cec} c(e^C_{\alpha}, d_{\beta})&=& k_1^2 \ [ \ V^{11}_1
V^{\alpha \beta}_2 + ( V_1 V_{UD})^{1
\beta}( V_2 V^{\dagger}_{UD})^{\alpha 1} ], \\
\label{ce} c(e_{\alpha}, d_{\beta}^C) &=& k^2_1  \ V^{11}_1
V^{\beta \alpha}_3 +  \ k_2^2 \
(V_4 V^{\dagger}_{UD} )^{\beta 1} ( V_1 V_{UD} V_4^{\dagger} V_3)^{1 \alpha},\\
\label{cnu} c(\nu_l, d_{\alpha}, d^C_{\beta})&=& k_1^2 \ ( V_1
V_{UD} )^{1 \alpha} ( V_3 V_{EN})^{\beta l} + \ k_2^2 \ V_4^{\beta
\alpha}( V_1 V_{UD}
V^{\dagger}_4 V_3 V_{EN})^{1l}, \ \ \ \alpha = 1 \ \text{or} \ \beta=1\\
\label{cnuc} c(\nu_l^C, d_{\alpha}, d^C_{\beta})&=& k_2^2 \ [ (
V_4 V^{\dagger}_{UD} )^{\beta
 1} ( U^{\dagger}_{EN} V_2)^{l \alpha }+ V^{\beta \alpha}_4
 (U^{\dagger}_{EN} V_2 V^{\dagger}_{UD})^{l1} ],  \ \ \ \alpha = 1 \ \text{or} \ \beta=1,
\end{eqnarray}
\end{subequations}
where $\alpha,\beta=1,2$. The physical origin of the relevant
terms is as follows. The terms proportional to $k_1(=g_{GUT}
M^{-1}_{V})$ are associated with the mediation of the superheavy
gauge fields $V=(X, Y)=(\overline{\bm{3}},\bm{2},5/6)$, where the
$X$ and $Y$ fields have electric charges $4/3$ and $1/3$,
respectively. This is the case of theories based on the SU(5)
gauge group. On the other hand, an exchange of
$V'=(X',Y')=(\overline{\bm{3}},\bm{2},-1/6)$ bosons yields the
terms proportional to $k_2(= g_{GUT} {M^{-1}_{V'}})$. In $SO(10)$
theories all these superheavy fields are present.

The relevant mixing matrices are $V_1= U_C^{\dagger} U$,
$V_2=E_C^{\dagger}D$, $V_3=D_C^{\dagger}E$, $V_4=D_C^{\dagger} D$,
$V_{UD}=U^{\dagger}D$, $V_{EN}=E^{\dagger}N$, $U_{EN}=
{E^C}^{\dagger} N^C$, and $V_{UD}=U^{\dagger}D=K_1 V_{CKM} K_2$,
where $K_1$ and $K_2$ are diagonal matrices containing three and
two phases, respectively. The leptonic mixing $V_{EN}=K_3
V^D_{PMNS} K_4$ in case of Dirac neutrino, or $V_{EN}=K_3
V^M_{PMNS}$ in the Majorana case. $V^D_{PMNS}$ and $V^M_{PMNS}$
are the leptonic mixing matrices at low energy in the Dirac and
Majorana case, respectively. Our convention for the
diagonalization of the up, down and charged lepton Yukawa matrices
is specified by $U^T_C Y_U U = Y_U^{\textrm{diag}}$, $ D^T_C Y_D D
= Y_D^{\textrm{diag}}$, and $E^T_C Y_E E = Y_E^{\textrm{diag}}$.

To establish the lower bound on the nucleon lifetime we first
specify the maximum value for all the coefficients listed above
for $SO(10)$ theory only. The $SU(5)$ case is well known and can
be reproduced by setting $k_2=0$ in the expressions below. The
upper bounds are
\begin{eqnarray}
c(e^C_{\alpha}, d_{\beta})_{SO(10)}& \leq & 2 \ k_1^2,\\
c(e_{\alpha}, d_{\beta}^C)_{SO(10)} & \leq & k^2_1  + k^2_2,\\
\sum_{l=1}^{3} c(\nu_l, d_{\alpha}, d^C_{\beta})^*_{SO(10)}
c(\nu_l, d_{\gamma}, d^C_{\delta})_{SO(10)} & \leq&  k_1^4 \delta^{\beta \delta} + k^4_2 + 2 k_1^2 k_2^2,\\
\sum_{l=1}^{3} c(\nu^C_l, d_{\alpha}, d^C_{\beta})^*_{SO(10)}
c(\nu^C_l, d_{\gamma}, d^C_{\delta})_{SO(10)} & \leq&  k_2^4 [ 3 +
\delta^{\alpha \delta}],
\end{eqnarray}
which translates into the following bounds on the amplitudes in
the case the neutrinos are Majorana particles:
\begin{eqnarray}
\Gamma(p \to \pi^+  \bar{\nu} ) & \leq &  \frac{m_p}{8 \pi
f^2_{\pi}} A_L^2 |\alpha|^2 (1 + D + F)^2
[k_1^2 + k_2^2]^2, \\
\Gamma(p \to K^+ \bar{\nu} ) & \leq & \frac{(m_p^2 - m_K^2)^2}{8
\pi f^2_{\pi} m_p^3} A_L^2
|\alpha|^2  \times \{ [\frac{2 m_p D}{3 m_B}]^2 + [ 1 + \frac{m_p}{3 m_B}(D + 3 F)]^2] \times \\
& & (k_1^2 + k_2^2 )^2  +  \frac{4 m_p D}{3 m_B} [ 1 + \frac{m_p}{3 m_B} (D + 3F)][ 2 k_1^2 k_2^2 + k_2^4 \},\\
\Gamma(p \to \pi^0  e^+_{\beta}) & \leq &  \frac{m_p}{16 \pi
f^2_{\pi}} A_L^2 |\alpha|^2 (1 + D + F)^2
[5 k_1^4 + 2 k_1^2 k_2^2 + k_2^4 ], \\
\Gamma(p \to K^0 e^+_{\beta}) & \leq & \frac{(m_p^2 - m_K^2)^2}{8
\pi f^2_{\pi} m_p^3} A_L^2 |\alpha|^2 [ 1 +
\frac{m_p}{m_B}(D-F)]^2 [ 5 k_1^2 + 2 k_1^2 k_2^2 + k_2^4 ].
\end{eqnarray}
Using these expressions it is easy to extract lower bounds on
lifetimes. In Table~\ref{tab:table1} we list all lower bounds for
the proton lifetime in $SU(5)$ and $SO(10)$ models. Again, we use
$m_p=938.3$\,MeV, $D=0.81$, $F=0.44$, $m_B=1150$\,MeV,
$f_{\pi}=139$\,MeV, $A_L=1.43$, and the most conservative value
$\alpha=0.003$\,GeV$^3$. In the case of $SO(10)$ models we set
$k_1=k_2$ for simplicity.
\begin{table}[h]
\caption{\label{tab:table1} Lower bounds for partial proton
lifetime in years for the Majorana neutrino case in units of
$M^4_V/\alpha_{GUT}^2$, where the mass of gauge bosons is taken to
be $10^{16}$\,GeV.}
\begin{ruledtabular}
\begin{tabular}{lccc}
Channel &  $\tau^{SU(5)}_p$ &  $\tau^{SO(10)}_p$\\
\hline
$p\rightarrow{{\pi }^+}\bar{\nu }$ &  $7.3 \times 10 ^{33}$ &   $1.8 \times 10 ^{33}$\\
$p\rightarrow{K^+}\bar{\nu}$ &   $17.4  \times 10 ^{33}$ &  $4.8  \times 10 ^{33}$\\
$p\rightarrow{\pi^0} e^+_{\beta}$  & $3.0 \times 10 ^{33}$  &  $1.8 \times 10 ^{33}$\\
$p\rightarrow{K^0} e^+_{\beta}$  &  $8.5 \times 10 ^{33}$ &  $5.3 \times 10 ^{33}$\\
\end{tabular}
\end{ruledtabular}
\end{table}
Note that lower bounds are well defined for the partial lifetimes
while upper bound is meaningful for the total lifetime only.

Finally, we can establish the theoretical bounds for the lifetime
of the proton in any given GUT. In what follows we use
\begin{equation}
10^{34}\,\text{years} \ \frac{M^4_V}{\alpha^2_{GUT}} < \tau_p <
10^{41}\,\text{years} \ \frac{M^4_V}{\alpha^2_{GUT}}.
\end{equation}
These bounds are useful since we can say something more specific
about the allowed values of $M_V$ or $\alpha_{GUT}$ or both. For
example, if we take $\alpha_{GUT}$ we can put lower limit on the
value of $M_V$ using experimental data on nucleon lifetime. Also,
given the value of $M_V$ and $\alpha_{GUT}$ we can set the limits
on the proton lifetime range within the given scenario.
\section{The two-loop running}
\label{third} We present the details on the two-loop running. In
order to maximize $M_V$ one needs extremely light $\Phi_b$. In any
given mass splitting scheme we thus set the mass of $\Phi_b$ at
$250$\,GeV which is just above the present experimental limit of
$242$\,GeV. Then, $\Sigma_3$ or $\Phi_a$ or both are allowed to
vary between $M_{\Phi_b}$ and $M_{GUT}$ in order to yield
unification and all other fields except for the SM ones are taken
to be at the GUT scale. The identification $M_V \equiv M_{GUT}$ is
justified through the inclusion of boundary conditions at the GUT
scale~\cite{Hall}
\begin{equation}
\alpha^{-1}_i|_{GUT}=\alpha^{-1}_{GUT}-\frac{\lambda_i}{12 \pi},
\qquad \{\lambda_1,\lambda_2,\lambda_3\}=\{5,3,2\},
\end{equation}
and the relevant two-loop equations for the running of the gauge
couplings take the well-known form
\begin{equation}
\mu \frac{\textrm{d}\,\alpha_i(\mu)}{\textrm{d}\,\mu}
=\frac{b_i}{2 \pi} \alpha^2_i(\mu)+\frac{1}{8 \pi^2}
\sum_{j=1}^{3} b_{ij} \alpha^2_i(\mu)\alpha_j(\mu).
\end{equation}
$b_i$ and $b_{ij}$ coefficients for the SM case for arbitrary $n$
are well-documented (see~\cite{Jones} for general formula). In
addition to those we have:
\begin{equation}
b_{i}^{\Sigma_3}= \left(\begin{array}{c}
   0 \\
   \frac{1}{3} \\
   0 \\
\end{array}\right),
\qquad
b_{i}^{\Phi_b}= \left(\begin{array}{c}
   \frac{1}{30} \\
   \frac{1}{2} \\
   \frac{1}{3} \\
\end{array}\right),
\qquad
b_{i}^{\Phi_a}= \left(\begin{array}{c}
   \frac{3}{5} \\
   \frac{2}{3} \\
   0 \\
\end{array}\right),
\end{equation}
\begin{equation}
b_{ij}^{\Sigma_3}=\left(
\begin{array}{ccc}
  0 & 0 & 0 \\
  0 & \frac{28}{3} & 0 \\
  0 & 0 & 0 \\
\end{array}\right),
\qquad
b_{ij}^{\Phi_b}=\left(
\begin{array}{ccc}
  \frac{1}{150} & \frac{3}{10} & \frac{8}{15} \\
  \frac{1}{10} & \frac{13}{2} & 8 \\
  \frac{1}{15} & 3 & \frac{22}{3} \\
\end{array}\right),
\qquad
b_{ij}^{\Phi_a}=\left(
\begin{array}{ccc}
  \frac{108}{25} & \frac{72}{5} & 0 \\
  \frac{24}{5} & \frac{56}{3} & 0 \\
  0 & 0 & 0 \\
\end{array}\right),
\end{equation}
which should be added to the SM ones at the appropriate particle
mass scale. The two-loop Yukawa coupling contribution to the
running of the gauge couplings (with the corresponding one-loop
running of Yukawas) is \textit{not}\/ included in order to make
meaningful comparison between $n=1$ and $n=2$ cases. (For the
$n=2$ case one vacuum expectation value of the light Higgs
doublets is arbitrary and needs to be specified in order to
extract fermion Yukawas for the running, i.e. $\tan \beta$
ambiguity. There is no such ambiguity present in the $n=1$ case
since the only low-energy vacuum expectation value is accurately
determined by electroweak precision measurements.)

\begin{figure}[h]
\begin{center}
\includegraphics[width=4.5in]{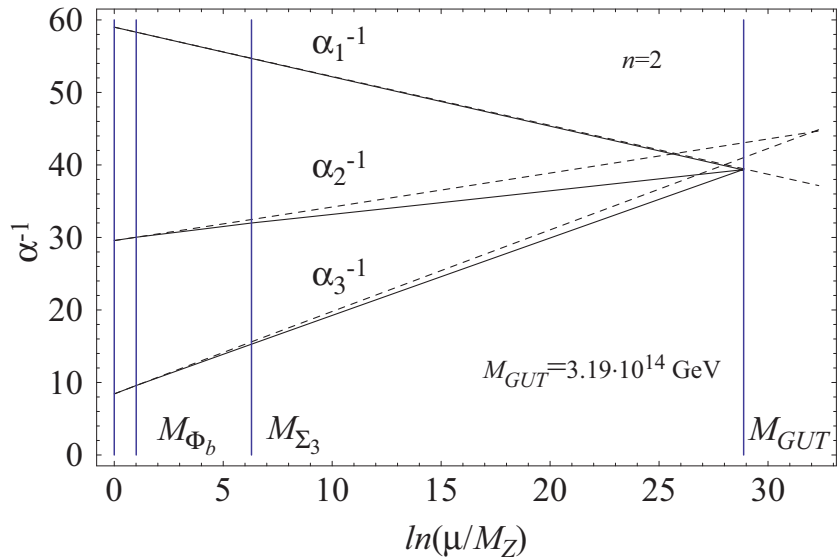}
\end{center}
\caption{\label{figure:6} Unification of the gauge couplings at
the two-loop level for central values of low-energy
observables~\cite{PDG2004}. The SM case with $n=2$ is presented by
dashed lines. Solid lines correspond to the $n=2$ scenario with
$\Phi_b$ and $\Sigma_3$ below the GUT scale. Vertical lines mark
the relevant scales: $M_Z$, $M_{\Phi_b}=250$\,GeV,
$M_{\Sigma_3}=4.95 \times 10^{4}$\,GeV and $M_{GUT}=3.19 \times
10^{14}$\,GeV.}
\end{figure}

\begin{figure}[h]
\begin{center}
\includegraphics[width=4.5in]{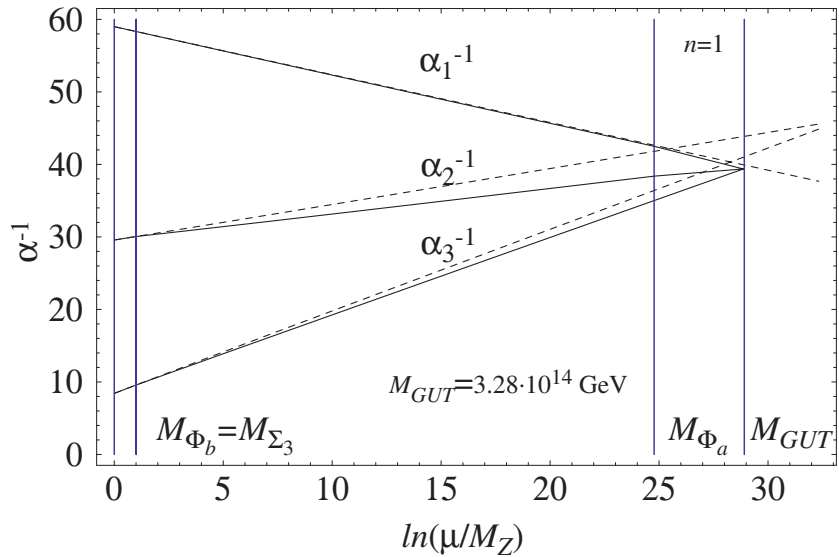}
\end{center}
\caption{\label{figure:7} Unification of the gauge couplings at
the two-loop level for central values of low-energy
observables~\cite{PDG2004}. The SM case with $n=1$ is presented by
dashed lines. Solid lines correspond to the $n=1$ scenario with
$\Phi_b$, $\Sigma_3$ and $\Phi_a$ below the GUT scale. Vertical
lines mark the relevant scales: $M_Z$,
$M_{\Phi_b}=M_{\Sigma_3}=250$\,GeV and $M_{GUT}=3.28 \times
10^{14}$\,GeV.}
\end{figure}

\begin{figure}[h]
\begin{center}
\includegraphics[width=4.5in]{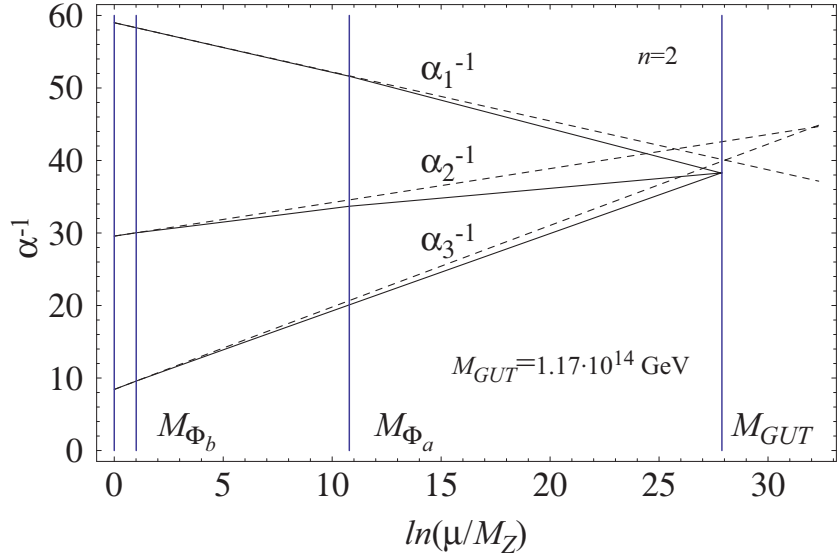}
\end{center}
\caption{\label{figure:8} Unification of the gauge couplings at
the two-loop level for central values of low-energy
observables~\cite{PDG2004}. The SM case with $n=2$ is presented by
dashed lines. Solid lines correspond to the $n=2$ scenario with
$\Phi_b$ and $\Phi_a$ below the GUT scale. Vertical lines mark the
relevant scales: $M_Z$, $M_{\Phi_b}=250$\,GeV, $M_{\Phi_a}=4.41
\times 10^{6}$\,GeV and $M_{GUT}=1.17 \times 10^{14}$\,GeV.}
\end{figure}

The outcome of the exact numerical unification is presented in
Figs.~\ref{figure:6},~\ref{figure:7} and~\ref{figure:8} .

The GUT scale as well as appropriate intermediate scales are
indicated on the plots. For example, in the $n=2$ scenario with
light $\Sigma_3$ in Fig.~\ref{figure:6} the GUT scale is close to
the one-loop results (see Fig.~\ref{figure:1} in particular) and
comes out to be $3.19 \times 10^{14}$\,GeV for central values of
coupling constants~\cite{PDG2004}. The $1\,\sigma$ departure
allows for the maximum value of $3.35 \times 10^{14}$\,GeV in that
case. Note that the $n=1$ case with light $\Sigma_3$ presented in
Fig.~\ref{figure:7} yields somewhat higher GUT scale. The reason
behind this trend is simple: the $\Phi_b$ contribution to $B_{12}$
is seven times that of the Higgs doublet but, at the same time,
they contribute the same to $B_{23}$. Thus, $\Phi_b$ is more
efficient in simultaneously improving the running and raising the
GUT scale than any extra Higgs doublets.

One might object that lightness of $\Sigma_3$, which requires
miraculous fine-tuning, makes the scenario with an extra $\bm{15}$
rather unattractive. Again, we do not insist on $\Sigma_3$ being
at intermediate scale. Note that the unification with almost the
same GUT scale as in the intermediate $\Sigma_3$ case is achieved
at the two-loop level in the scenario where $\Phi_a$ is at
intermediate and $\Sigma_3$ is at the GUT scale. See
Figs.~\ref{figure:4} and~\ref{figure:8} for the $n=1$ and $n=2$
cases, respectively. As discussed in the text, intermediate scale
for $\Phi_a$ would require either small Yukawas for Majorana
neutrinos or small $c_3$. In the latter case, the novel
contributions towards proton decay would be automatically
suppressed. The former case could be probed if and when the
leptoquarks are detected since some of the rare processes
involving neutrinos would be significantly suppressed compared
with the charged lepton ones. We note that in the scenario with
intermediate $\Phi_a$ the GUT scale grows with the number of light
Higgs doublets in contrast to the case when $\Sigma_3$ is at the
intermediate scale. Again, the reason is that the $B_{12}$
coefficient of $\Phi_a$ is the same as the appropriate coefficient
of the Higgs doublet but its contribution to $B_{23}$ is four
times bigger. Thus, the very efficiency of $\Phi_a$ in improving
unification makes its impact on $M_{GUT}$ rather small.


\end{document}